\documentstyle[aps,twocolumn,prl]{revtex}

\input epsf
\def\plotone#1{\centering \leavevmode
\epsfxsize= 1.\columnwidth \epsfbox{#1}}
\def\plottwo#1{\centering \leavevmode
\epsfxsize= 0.75\columnwidth \epsfbox{#1}}

\def\bff{}
\def\etal{{et al.}}
\def\apjl{Astrophys. J. Lett.}
\def\mnras{Mon.Not.Roy.As.Soc.}

\def\be{\begin{equation}}
\def\ee{\end{equation}}
\def\bea{\begin{eqnarray}}
\def\eea{\end{eqnarray}}

\def\cmm2{{\,\rm cm^{-2}}}
\def\cm2{{\,{\rm cm}^2}}
\def\cmm3{{\,{\rm cm}^{-3}}}
\def\gcmm3{{\,{\rm g\,cm^{-3}}}}

\def\fun#1#2{\lower3.6pt\vbox{\baselineskip0pt\lineskip.9pt
  \ialign{$\mathsurround=0pt#1\hfil##\hfil$\crcr#2\crcr\sim\crcr}}}

\def\etal{{et al.}}

\def\p3m{P$^3$M}

\def\fun#1#2{\lower3.6pt\vbox{\baselineskip0pt\lineskip.9pt
  \ialign{$\mathsurround=0pt#1\hfil##\hfil$\crcr#2\crcr\sim\crcr}}}

\begin{document}
\twocolumn[\hsize\textwidth\columnwidth\hsize\csname @twocolumnfalse\endcsname
\draft
\title{Dark Energy and the CMB}
\author{Scott\ Dodelson$^1$ and Lloyd\ Knox$^2$}
\address{$^1$ NASA/Fermilab Astrophysics Center\\
Fermi National Accelerator Laboratory, Batavia, IL 60510, USA}
\address{$^2$ Department of Astronomy and Astrophysics\\
University of Chicago, 5640 S. Ellis Ave., Chicago, IL 60637, USA}
\date{\today}
\maketitle

\begin{abstract}
  We find that current Cosmic Microwave Background (CMB) anisotropy
  data strongly constrain the mean spatial
  curvature of the Universe to be near zero, or, equivalently, the
  total energy density to be near critical---as predicted by inflation.  
  This result is robust to
  editing of data sets, and variation of other cosmological parameters
  (totaling seven, including a cosmological constant).  Other lines of
  argument indicate that the energy density of non-relativistic matter
  is much less than critical.  Together, these results are evidence,
  independent of supernovae data, for dark energy in the Universe.
\end{abstract}
] 
{\parindent0pt\it Introduction.} 
Cosmologists have long realized that there is more to the Universe
than meets the eye. A wide variety of evidence points to
the existence of dark matter in the Universe, matter which 
cannot be seen, but which can be indirectly detected
by its contribution to the gravitational field. As observations
have improved, the phenomenology of the ``dark sector'' has become richer.
While dark matter was originally posited to explain what would
otherwise be excessively attractive gravity, dark energy explains
the accelerating expansion---an apparently repulsive gravitational
effect. The most well-known argument for
this additional dark component 
is based on inferences of the luminosity distances
to high--$z$ supernovae\cite{snIa}.  The anomalously large distances indicate
that the Universe was expanding more slowly in the past
than it is now; i.e., the expansion rate is
accelerating.  Acceleration only occurs if the bulk pressure
is negative, and this could only be due to a previously
undetected component.

Here we argue for dark energy based on another
gravitational effect: its influence on the mean spatial curvature.
This argument \cite{believe} does not rely on the supernovae 
observations and therefore avoids the systematic uncertainties
in the inferred luminosity distances.
It is based on a lower limit to the
total density, and a smaller upper limit on the density of non-relativistic
matter.  The lower limit comes from measurements of the anisotropy of
the cosmic microwave background (CMB) whose statistical properties
depend on the mean spatial curvature\cite{kamionkowski94},
which in turn depends on the mean total density.  
We find that the CMB strongly indicates that $\Omega > 0.4$,
where $\Omega$ is the ratio of the total mean density to the critical
density (that for which the mean curvature would be zero).  
Upper limits to the density of non-relativistic matter come from
a variety of sources which quite firmly indicate $\Omega_m < 0.4$.

The CMB sensitivity to curvature is due to the dependence on curvature
of the angular extent of objects of known size, at known redshifts.
CMB photons that are penetrating our galaxy today, were emitted from a
thin shell at a redshift of $z \simeq 1100$ (called the
``last-scattering surface'') during the transition from an ionized
plasma to a neutral medium.  The ``object'' of known size at known
redshift is the sound-horizon of the plasma at the epoch of
last-scattering.  Its observational signature is the location of a
series of peaks in the angular power spectrum of the CMB.

One must be careful about using current CMB data to determine $\Omega$ or
any other cosmological parameters for several reasons.  First, these
are very difficult experiments, and the data sets they produce have low
signal-to-noise ratios and limited frequency ranges, complicating the
detection of systematic errors.  Use of different calibration
standards further increases the risk of underestimated systematic
error.  To counter these problems, 
we examine the robustness of our results to editing of
data sets, and check that the distribution of model residuals is
consistent with the stated measurement uncertainties.

Second, the CMB angular power spectra depend on a number of parameters
other than the curvature.  To some
degree, a change in curvature can be mimicked by changes in other
parameters.  We therefore vary six parameters besides the
curvature, placing mild prior constraints on some of these
so as not to explore unrealistic regions of the parameter space.

Finally, existing data are insufficient to firmly establish the paradigm for
structure formation which we have assumed:
structure grew via gravitational instability from primordial 
adiabatic perturbations.  
Our conclusions {\it depend} on this assumption. At present,
this counts as a possible source of systematic error.
Fortunately, future CMB data will verify (or refute) the paradigm
and will also allow for the determination of $\Omega$ with greatly
reduced model dependence\cite{huwhite}. 

{\parindent0pt\it The data.}  Present data are already so abundant that
it must be compressed before it can serve as the basis for a multi-dimensional
parameter search.  
Fortunately, all data sets have been compressed to constraints
on the angular power spectrum, $C_l \equiv 2\pi \int C(\theta) 
P_l(\cos\theta)d(\cos\theta)$ where $C(\theta)$ is the
correlation function.  Because of the tremendous reduction in
the size of the data sets, this 
data compression is called ``radical compression'' \cite{BJKII}.

Here we use the radically compressed data from
{\tt http://www.cita.utoronto.ca/\~knox/radical.html}.
In this compilation the non-Gaussianity of the
power spectrum uncertainties has been characterized 
for a number of experiments with a lot of the weight
(including all those plotted with large symbols in Fig. 1);
assuming Gaussianity leads to
biases\cite{BJKII}.  

\begin{figure}[htbp]
  \plotone{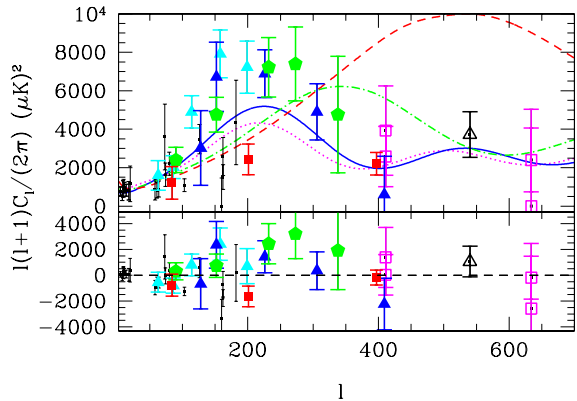} 
    \caption{Constraints on the angular power spectrum:  
those marked with large symbols are 
TOCO (filled triangles) \protect\cite{TOCO}, 
CAT (open squares) \protect\cite{CAT}, 
SK (pentagons) \protect\cite{SK}, 
OVRO5M (open triangle) \protect\cite{OVRO5M} and 
MSAM (filled squares) \protect\cite{MSAM}. 
The model curves are standard COBE-normalized CDM (dotted), the best-fit
$\Omega=1$ model (solid), the best-fit $\Omega = 0.4$ model
(dot-dashed), and the best-fit $\Omega = 0.2$ model (dashed). The lower
panel shows residuals of the best-fit $\Omega=1$ model.
}
\label{fig:dkfig1}
\end{figure}

{\parindent0pt\it The Search Method.}  We search over a 
seven-dimensional parameter space specified by
$\Omega$, $\Omega_b h^2$, $\Omega_{\rm cdm} h^2$,
$\Omega_{\Lambda}h^2$, $\tau$, $n_s$ and $C_{10}$, where $\Omega_i =
\rho_i/\rho_c$ and $i=b,{\rm cdm},\Lambda$ is for baryons, cold dark
matter and a cosmological constant respectively, 
$\rho_c \equiv 3H_0^2/(8\pi G)$ is the critical density, $\tau$ is the optical
depth to Thomson scattering, $n_s$ is the power-law index of the
primordial matter power spectrum, and $C_{10}$ serves as the
normalization parameter.  The Hubble
constant, $H_0 \equiv {100h\,{\rm km\,sec^{-1}\,Mpc^{-1}}}$,
is a dependent variable in this space, due to the sum rule:
$\Omega_\Lambda+\Omega_b+\Omega_{\rm cdm} = \Omega$.  
Note that, for specificity and simplicity, we have chosen the dark energy
to be a cosmological constant; other choices (e.g., qunitessence\cite{quintessence}) would not significantly affect 
our curvature constraints.

For each value of $\Omega$ we vary the 23 other parameters (six cosmological
and 17 calibration---one for each experiment)
to find the
minimum value of $\chi^2 = \chi^2_d + \chi^2_p$. Here 
$\chi^2_d$ is the offset log-normal form explicitly
given in Eq. 39-43 of \cite{BJKII}, which was shown to be
a good approximation to the log of the likelihood function.
Information from non-CMB observations is included as a prior contribution,
$\chi^2_p$.  Unless otherwise stated, we assume
that $h = 0.65 \pm 0.1$ (a reasonable interpretation of
several measurements\cite{hubconst}) and $\Omega_b h^2 = 0.019 \pm 0.003$
(from \cite{BNTT99} but with a 40\% increase in their uncertainty).
We use the Levenberg-Marquardt method to find the minimum value of $\chi^2$
for each value of $\Omega$.
We stop the hunt when the new $\chi^2$ is within $0.1$
of the old value. We tested this method on simulated data and recovered
the correct results.  

The likelihood of the best-fit model, ${\cal L}(\Omega)$ 
is proportional to $\exp(-\chi^2/2)$.  Ideally we would marginalize
over the non-$\Omega$ parameters rather than maximizing over them.
However, we note that in the limit that the likelihood is Gaussian,
these two procedures are equivalent.  More generally, in order for
marginalization to give qualitatively different answers there would
have to be, with decreasing $\Omega$, a very rapid increase in the
volume of parameter space in the non-$\Omega$ direction with $\chi^2$'s
comparable to the minimum $\chi^2$.  Inspection of the Fisher matrix
leads us to believe this is not the case.  

{\parindent0pt\it The Results.} Our main results are shown in Fig. 2:
the relative likelihood ($\propto \exp(-\chi^2/2)$) of 
the different values of $\Omega$.  
Including all the data, the best-fit (minimum
$\chi^2$) $\Omega = 1$ model is $2\times 10^7$ times more
probable than the best-fit $\Omega = 0.4$ model. $\Omega < 0.7$
is ruled out at the $95\%$ confidence level.

\begin{figure}[htbp]
  \plotone{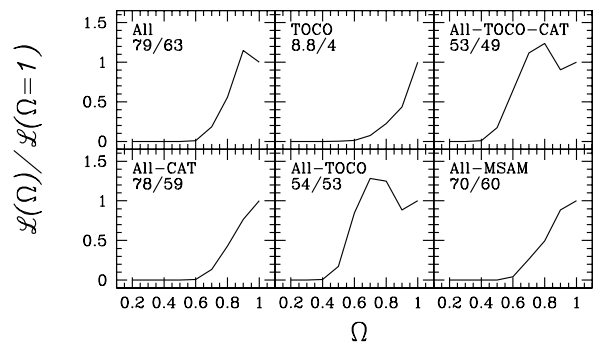} 
    \caption{Relative likelihood of $\Omega$ and $\chi^2$ over the 
degrees of freedom (for $\Omega=1$) for different collections
of data sets. }
\label{fig:dkfig2}
\end{figure}

To test the robustness of this result, we edited
out single data sets suspected of providing the most weight. Most of these
editings produced little change. Only the omission of
TOCO changes things substantially, and even then, the best-fit 
$\Omega = 1$ model is $150$ times more probable than the best-fit 
$\Omega = 0.4$ model.
We also edited pairs of data sets:  for no CAT and TOCO, no MSAM and CAT,
and no MSAM and TOCO, we find $\Omega=1$ to be $120$, $2.5\times 10^6$ and
$8$ times more likely than $\Omega=0.4$.  Also shown, as measures
of goodness-of-fit, are $\chi^2$ and the degrees of freedom.  
The $\chi^2$ value for the ``All'' case  is a bit high, 
but one expects even higher ones over $8\%$
of the time, so there is no strong evidence for inconsistencies in
the data.  As further indication of the robustness of the result,
one can see from the ``TOCO'' panel of Fig. 2 that it persists
even when all but a single data set is removed.

For the ``All'' case, the best-fit $\Omega=1$ model has $\Omega_b h^2
= 0.019$, $h=0.65$, $\Omega_\Lambda=0.69$, $\tau =0.17$ and $n=1.12$
and is plotted in Fig. 1.  There are degeneracies among these
parameters though and none of them is strongly constrained on its own.
For example, an equivalently good fit (to just the CMB data) is given
by the following model with no tilt or reionization: $\Omega=1$,
$\Omega_b h^2=0.021$, $h=0.65$, $\Omega_\Lambda=0.65$, $\tau=0$ and
$n=1$.

We also covered the $\Omega_m$, $\Omega_{\Lambda}$ plane, at each
point finding the minimum $\chi^2$ possible with variation of the
remaining 5 parameters. Figure~3 shows the $\Delta \chi^2=$ 1, 4 and 9
contours in this plane (which, for a Gaussian, correspond to
the 40\%, 87\% and 99\% confidence regions, respectively).

\begin{figure}[b]
\plottwo{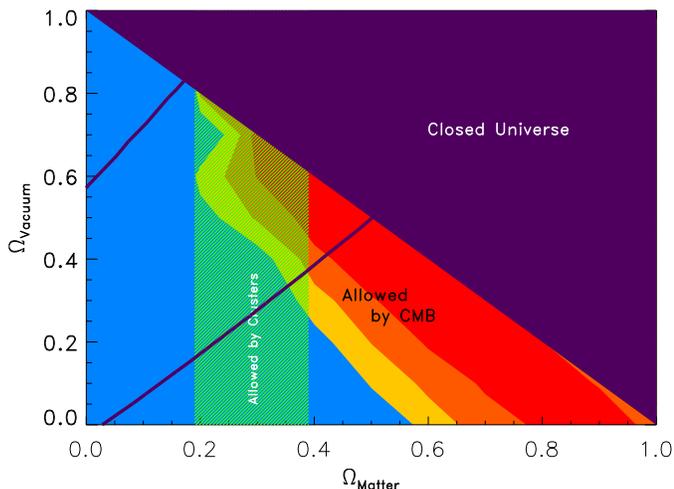} 
\caption{Likelihood contours in the $\Omega_m$, $\Omega_{\Lambda }$
plane.  Contours show $\Delta \chi^2 = 1,4,9$.
In green is the 95\% confidence region for $\Omega_m$ from cluster baryon
fraction determinations. The two solid lines encompass the
allowed $2-\sigma$ region from super nova data\protect\cite{snIa}.}
\label{fig:dkfig3}
\end{figure}

{\parindent0pt\it Discussion.}
Figure 3 also shows constraints on $\Omega_m$ from clusters.
Although constraints on $\Omega_m$
arise from a variety of techniques (for reviews see \cite{omegam}) perhaps the most reliable are those based on the
determination of the ratio of baryonic matter to dark matter in
clusters of galaxies 
\cite{white93,clustersplussims,clustersonly,mohr99,grego99}.
With the assumption that the cluster ratio is the mean ratio
(reasonable due to the large size of the
clusters)\cite{white93,clustersplussims,mohr99,clustersimsonly}, 
and the baryonic mean density from
nucleosynthesis, one can constrain the range of allowable values of
$\Omega_m$.  Since only the baryonic intracluster {\it gas} is
detected, the upper limits on $\Omega_m$ from this method are better
understood than the lower limits.  Mohr et al. \cite{mohr99} find,
from a sample of 27 X-ray clusters, that (including corrections for
clumping and depletion of the gas) $\Omega_m < (0.32\pm
0.03)/\sqrt{h/0.65}$. Including the Hubble constant uncertainty
($h=0.65 \pm 0.1$) this becomes $\Omega_m < 0.32 \pm 0.05$.  Assuming
10\% of the baryons to be in galaxies as opposed to the gas, as
estimated by \cite{white93}, we find $\Omega_m = 0.29 \pm 0.05$.
Results from observations of the Sunyaev-Zeldovich effect in
clusters are consistent, though
less restrictive: $\Omega_m = 0.31 \pm 0.1$ \cite{grego99}.  
Most other methods
(those that do not rely on the cluster baryon fraction) generally
result in formally stronger upper limits to $\Omega_m$.  This increases our
confidence in the Mohr et al. $\Omega_m$ upper limit, but we do not
quote these stronger constraints due to our concerns that they are
affected by systematic uncertainties that are more difficult to
quantify than those in the baryon fraction method.

There have been a number of other analyses\cite{previouswork} of
CMB anisotropy data which generally obtained weaker constraints on
$\Omega$\cite{endnote}. 
There are technical differences between our work and
previous work: we account for the non-Gaussianity of the likelihood
function, allow for calibration uncertainties, place ``sanity'' priors
on the Hubble constant and the baryon density, and vary six parameters
in addition to the curvature. Also, much of the strength of our
argument comes from data reported within the last year.  

The verdict
from the CMB is now in. It does not depend on any one, or even any
two, experiments. It clearly
points towards a flat Universe and, together with cluster data,
strongly indicates the existence of dark energy. These
conclusions are consistent with, and independent of, the supernovae
results. The completely different sets of systematic uncertainties in the two
arguments further strengthen the case.  Other constraints in the 
$\Omega_m, \Omega_\Lambda$ plane were recently obtained \cite{zehavi}
by combining cosmic flow data with supernovae observations.

We have neglected several data sets, all of which, if included, would
only strengthen our conclusions.  Two of these are PythonV\cite{PyV}
and Viper\cite{viper}.  PythonV and
Viper together trace out a peak with centroid near $l = 200$, and a
significant drop in power by $l=400$.  They have not been included
because of the strong correlations in the existing reductions of the
data; a new reduction with all correlations specified will soon be
available for PythonV.  

Any model without a drop in power from $l=200$ to $l=400$
has difficulties
agreeing with all the data.
Models fitting this
description include the adiabatic models considered here with
$\Omega < 0.4$ and also topological defect models, whose breadth
is a consequence of the loss of the coherent peak structure \cite{coherence}.

We have been concentrating on implications of
the peak location, but the height is also of interest.
With fixed $h$, it is additional evidence for
low $\Omega_m$.  The lower $\Omega_m h^2$, the later the
transition from a radiation-dominated Universe to a matter-dominated
Universe and the larger the early ISW effect, which contributes
in the region of the first peak \cite{husug}.  For flat models,
the best fit is at $\Omega_m=0.4$ with $\Omega_m=1$ four times less likely.

{\parindent0pt\it Conclusions.}
We have shown that $\Omega=1$ is strongly favored over $\Omega =
0.4$.  This result is interesting for two reasons.  First, $\Omega =1$ is a
prediction of the simplest models of inflation. Second, together
with the constraint $\Omega_m < 0.4$, it is
evidence for dark energy. 

The CMB can say little about the nature of the dark energy. A
cosmological constant fits the current data, but then so would many of
the other forms of dark energy proposed over the past few years.
Generation and exploration of new theoretical ideas as to the nature
of this dark energy is clearly warranted.

Measurements of CMB anisotropy have already delivered on their promise
to provide new clues towards an improved understanding of cosmological
structure formation and fundamental physics.  We look forward to
greater clarification of the dark energy problem, as well as possibly
new surprises, from improved CMB anisotropy measurements in the near
future.

\acknowledgements 
We are grateful to J. Mohr for useful conversations and A. Jaffe
for supplyings us with the SN data.  
We used CMBFAST many, many times\cite{cmbfast}.  
SD is supported by the DOE and by NASA Grant NAG 5-7092.
LK is supported by the DOE, NASA grant NAG5-7986 and NSF grant OPP-8920223.

\end{document}